\documentclass[twocolumn, secnumarabic, amssymb, nobibnotes, prl, superscriptaddress]{revtex4}

\usepackage{graphicx} 
\usepackage{amsmath}
\usepackage{float}
\usepackage{color}
\usepackage[english]{babel}
\usepackage{array}
\usepackage[T1]{fontenc}
\usepackage[usenames,dvipsnames]{xcolor}
\usepackage{setspace}
\usepackage{hyperref}
\usepackage{bm}
\usepackage{lipsum} 
\usepackage{tikz}
\usepackage{stmaryrd}
\makeatletter
\renewcommand*{\fnum@figure}{{\normalfont \small{FIG.}~\thefigure}}
\makeatother

\begin{document}

\title{Dissipation indicates memory formation in driven disordered systems}

\author{Dor Shohat}
\affiliation{Department of Condensed Matter, School of Physics and Astronomy, Tel Aviv University, Tel Aviv 69978, Israel}
\affiliation{Center for Physics and Chemistry of Living Systems, Tel Aviv University, Tel Aviv 69978, Israel}
\author{Yoav Lahini}
\affiliation {Department of Condensed Matter, School of Physics and Astronomy, Tel Aviv University, Tel Aviv 69978, Israel}
\affiliation{Center for Physics and Chemistry of Living Systems, Tel Aviv University, Tel Aviv 69978, Israel}

\begin{abstract} Disordered and amorphous materials often retain memories of perturbations they have experienced since preparation. Studying such memories is a gateway to understanding this challenging class of systems. However, it often requires the ability to measure local structural changes in response to external drives. Here, we show that dissipation is a generic macroscopic indicator of the memory of the largest perturbation. Through experiments in crumpled sheets under cyclic drive, we show that dissipation transiently increases when first surpassing the largest perturbation due to irreversible structural changes with unique statistics. This finding is used to devise novel memory readout protocols based on global observables only. The general applicability of this approach is demonstrated by revealing a similar memory effect in a three-dimensional amorphous solid.   
\end{abstract}

\maketitle
The properties of far-from-equilibrium, disordered systems depend strongly on their history - their preparation protocol, initial conditions, and past external perturbations. As a result, their structural properties and macroscopic response vary between realizations, making them hard to predict or reproduce. Nevertheless, this strong history dependence offers a fruitful approach to study these systems - by considering the memories they hold of their past \cite{keim2019memory}. Memories constitute robust and reproducible phenomena in an otherwise irregular system. Thus, they offer a road map for exploring the system's complex energy landscape, which is high-dimensional, rough, and contains many metastable minima \cite{mungan2019networks,regev2021topology,lindeman2021multiple,bense2021complex}. In addition, as a wide range of disordered systems exhibit similar memory phenomena, studying them may help reveal common motifs and uncover general principles \cite{keim2019memory}.

Perhaps the most ubiquitous and robust form of memory, displayed by many disordered systems, is the memory of the largest perturbation \cite{keim2019memory}. Systems as diverse as disordered packings and amorphous solids \cite{keim2011generic,keim2020global}, particle suspensions and emulsions \cite{corte2008random}, biological tissues and gels \cite{munster2013strain,majumdar2018mechanical}, magnetic systems \cite{gilbert2015direct}, and crumpled sheets \cite{matan2002crumpling,shohat2022memory}, all carry distinct signatures of the largest strain, shear, load, or magnetic field applied to them. Namely, their structure or spatial configuration is generically set by the largest applied perturbation. In many cases, this memory can be consolidated using periodic driving, during which the system adapts to the drive until reaching an approximate or exact periodic response \cite{keim2019memory}. At this state, irreversible configurational changes only occur if the amplitude of the training drive is exceeded, as suggested by Corté \textit{et al.} \cite{corte2008random}.

Keim and Nagel \cite{keim2011generic} offered a framework for quantifying this memory using rigorous readout protocols. Namely, one can infer whether a cycle exceeded the largest perturbation by monitoring irreversible structural changes in the material under cyclic driving. This approach has proven useful for studying cyclic memories in various amorphous solids, illuminating important features of the energy landscape which give rise to memory formation \cite{fiocco2014encoding,adhikari2018memory,mukherji2019strength,keim2022mechanical,lindeman2021multiple}, while exploring the roles of noise \cite{keim2011generic,paulsen2014multiple,paulsen2019minimal}, dimensionality \cite{schwen2020embedding}, and inter-particle forces \cite{benson2021memory,chattopadhyay2022inter}.

However, these memory readouts require access to local degrees of freedom or the microscopic structure, which is not always possible due to geometry, dimensionality, or experimental constraints. While evidence for changes in bulk moduli upon surpassing the largest perturbation has been reported \cite{matan2002crumpling, paulsen2014multiple,mukherji2019strength,chattopadhyay2022inter}, this signal is often weak, and the relation between local and global signatures of memory formation is not well understood. In this letter, we propose that dissipation, measured using either local or global observables, is a generic and robust indicator for memory formation in disordered systems under cyclic drive. We show that when the largest perturbation is first exceeded, irreversible structural changes, accompanied by excess energy dissipation, result in unique, globally measurable signals. These are used to formulate global memory readout protocols. Our results link structural memory signals to their macroscopic signatures, enabling the study of memory in systems where the internal structure is inaccessible.

To correlate local and global memory signatures, we study cyclic memory formation in crumpled sheets, an accessible table-top amorphous system recently shown to exhibit memory of largest strain \cite{shohat2022memory}. We show that this memory can be retrieved in three different ways: by monitoring irreversible local changes in the sheet's geometry in response to cyclic drive, and from two global signals of excess energy dissipation. The first is inferred from the hysteresis in the sheet's stress-strain curves; the hysteresis area, representing the work dissipated over a cycle, increases transiently when the largest applied strain is surpassed. The second signal can be obtained by tracking the excess acoustic energy emitted by the sheet as it deforms. Finally, to demonstrate the generality of our approach, we show that global dissipation signals can be used to extract encoded memories in a less approachable system: steel wool, a three-dimensional frictional amorphous solid.

\textit{Experimental system - } 
Our starting point is the recently observed memory of largest strain in thin sheets of Mylar that have been crumpled many times \cite{shohat2022memory}. It was shown that when a crumpled sheet is deformed, its shape changes via a series of abrupt local mechanical instabilities. In each of these sudden events a localized region of the sheet snaps between two possible configurations \cite{lechenault2015generic,andrade2019foldable}. Every such instability has two global signatures: a stress jump in the global mechanical response, and a pulsed acoustic emission \cite{houle1996acoustic,kramer1996universal}. Under cyclic strain and without any additional crumpling, both the shape of the sheet and its hysteretic mechanical response were found to be set by the largest strain the sheet has experienced. In particular, it was shown that upon surpassing the largest strain applied, irreversible local instabilities occur - permanently changing both the sheet's geometry and its mechanical response.

We now turn to quantify this memory using three different signals, leveraging the macroscopic nature of crumpled sheets which allows tracking and correlating responses across scales, from the local structure to macroscopic observables. To measure the response, we use the custom mechanical testers described in Ref. \cite{shohat2022memory}. Here, an unfolded crumpled sheet is strained using a linear stage by setting its displacement $\Delta$, while the resistance force $F$ exerted by the strained sheet is measured using a load cell. The measured force-displacement curves can be correlated with structural changes in the sheet, obtained using 3D imaging. In addition, a microphone placed near the sheet records the acoustic emission.

\textit{Local structural signal - }
We start by showing that, similarly to other driven disordered systems \cite{keim2011generic,paulsen2014multiple,fiocco2014encoding}, the memory of largest strain in crumpled sheets can be extracted by tracking changes in its structure. First, memory of largest strain is encoded into a crumpled sheet by periodically straining it, i.e., by cycling the stage over a constant displacement interval $\left[\Delta_0,\Delta_{train}\right]$ and reaching a limit cycle in the force-displacement curves. From this point onward, the sheet returns to the same geometrical configuration at the end of additional cycles over the same interval \cite{shohat2022memory}. Namely, all the geometric instabilities that occur during the compression part of the limit cycle are reversed during extension. Moreover, cycling the stage over $\left[\Delta_0,\Delta_{1}\right]$ for any $\Delta_{1}<\Delta_{train}$, does not change the spatial configuration of the sheet at the end of the cycle, $\Delta_0$. In contrast, for any $\Delta_{1}>\Delta_{train}$, irreversible transformations occur and the shape of the sheet at the end of the cycle changes.

\begin{figure}
\includegraphics[width=0.45\textwidth]{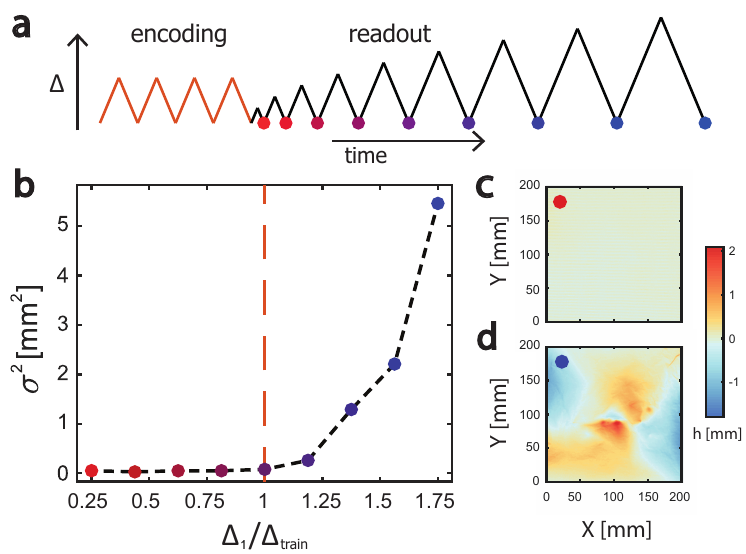}
\vspace*{-0.2cm}
\caption{\textbf{Local structural signal - } (\textbf{a}) The strain protocols used to encode and retrieve the memory. To encode a memory the stage is oscillated over a fixed amplitude $\Delta_{train}$. For readout the amplitude is slowly increased, and a 3D scan is performed after each cycle (marked by circles); (\textbf{b}) Mean square error $\sigma^2$ between 3D scans and a reference scan taken after encoding, vs normalized readout amplitude $\Delta_{1} / \Delta_{train}$, for $\Delta_{train}=80\,$mm. Above $\Delta_{train}$, $\sigma^2$ exhibits a sharp rise, indicating the encoded memory; (\textbf{c-d}) 3D scan differences for $\Delta_{1} / \Delta_{train}=0.25$ and $1.75 $ respectively.}
\label{fig_local}
\end{figure}

The geometric differences between states of the sheet, measured at the same reference displacement, can be used to retrieve the encoded memory. To this end, we obtain 3D scans of the sheet at $\Delta_0$, at different stages of the experiment. These scans are then registered on the plane formed by the corners of the sheet $h(x,y)$. To compare states, we subtract the height maps and calculate the mean square error (MSE)
\begin{equation}
\sigma^2=\frac{1}{A}\int\left(h_{1}-h_{0}\right)^2\,dA
\end{equation}
over the area of the sheet $A$. The $\sigma^{2}$ measure can be used to retrieve an encoded memory. After encoding the memory of $\Delta_{train}$, we obtain a 3D scan at $\Delta_{0}$, denoted $h_{0}$. The strain is then cycled over $\left[\Delta_{0},\Delta_{1}\right]$, gradually increasing the maximal strain $\Delta_{1}$, covering a large range of amplitudes (Fig. \ref{fig_local}a). After each full cycle, we obtain the 3D topography $h_{1}$ and calculate $\sigma^2$ compared to the initial state. The results, presented in Fig. \ref{fig_local}b, show that below the trained amplitude the MSE is close to zero, while above it the memory signal rises sharply.

\textit{Global mechanical signals - }
By examining the global strain-stress response curves, we find that they also hold distinct signatures of memory formation. Different features of these curves allow discerning between a transitory, irreversible response cycle, and a limit cycle reached after training. Since transient cycles only occur above the largest applied strain \cite{shohat2022memory}, we harness these differences to formulate global readout signals.

\begin{figure}
\includegraphics[width=0.45\textwidth]{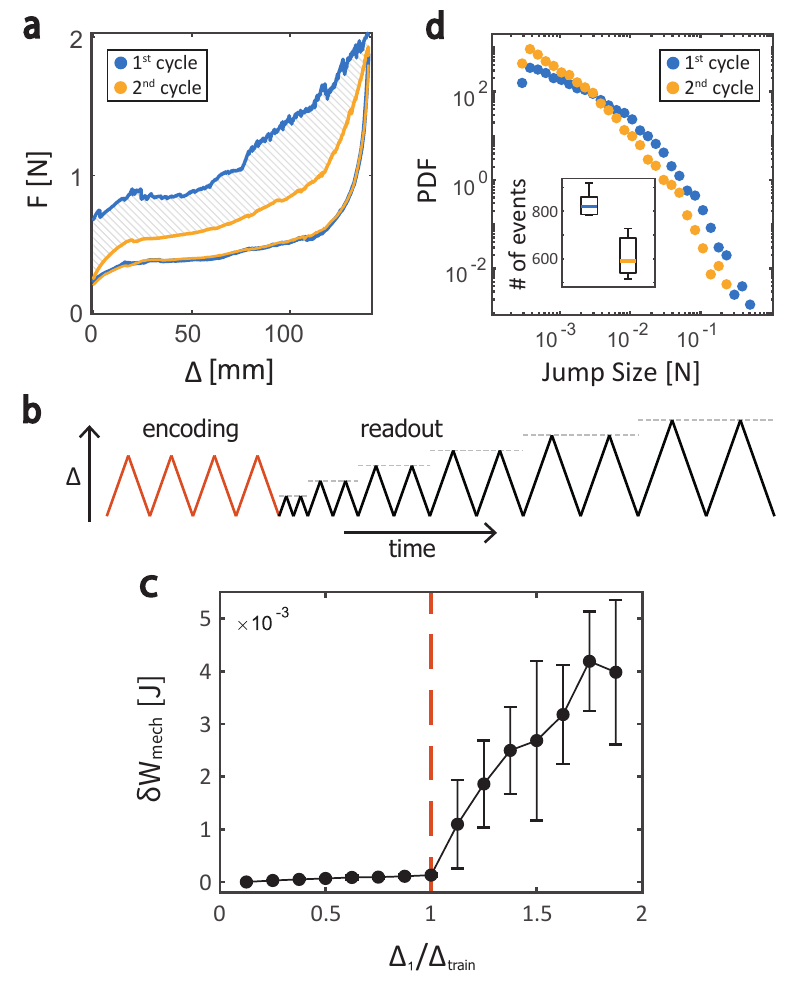}
\vspace*{-0.2cm}
\caption{\textbf{Global mechanical signals - } (\textbf{a}) Force vs. stage displacement $\Delta$ for two consecutive cycles. The first, transitory cycle dissipates more energy than the second. The excess dissipation is indicated by the marked area; (\textbf{b}) $\Delta$ over time for the global readout protocol, where each amplitude is cycled twice; (\textbf{c}) Excess dissipation between the two cycles $\delta W_{mech}$ as a function of cycle amplitude $\Delta_{1} / \Delta_{train}$, for $\Delta_{train}=40\,$mm, averaged over $N=8$ trials. The sharp rise above $1$ signals the memory. (\textbf{d}) PDF of the force jump sizes, comparing transitory trajectories and limit cycles. Transients are characterized by larger drops, and a higher number of events (inset); }
\label{fig_forces}
\end{figure}

We first consider an untrained crumpled sheet, with no encoded memory. Performing several cycles over a constant strain interval, results in one prominent transient cycle, followed by a quick convergence to a limit cycle (Fig. \ref{fig_forces}a). The transitory cycle exhibits excess energy dissipation. Namely, the overall work consumed during this cycle is much larger than the work consumed during a second cycle over the same interval or a limit cycle. This excess dissipation is highlighted in Fig. \ref{fig_forces}a. This also holds for transients that occur after memory encoding. Thus, we can devise a dissipation based memory readout protocol, relying purely on global measurements. 

Following an identical encoding protocol, we cycle the strain over $\left[\Delta_{0},\Delta_{1}\right]$, while gradually increasing the upper limit. This time, however, \textit{two} oscillations are performed for each $\Delta_{1}$, as shown in Fig. \ref{fig_forces}b. By subtracting the hysteresis loop areas of the two cycles
\begin{equation}
\delta W_{mech}=\oint\left(F_{1}-F_{2}\right)d\Delta
\end{equation}
we measure the excess dissipation, and obtain a global memory signal \cite{footnote_cycles}. For $\Delta_{1}<\Delta_{train}$ there are no transients, the system immediately falls into a limit cycle, and the areas of the two consecutive loops are identical up to material creep \cite{jules2020plasticity,supplementary} \nocite{regev2013onset,feldman2011hilbert}
. However, for $\Delta_{1}>\Delta_{train}$ the first cycle is transitory and dissipates energy towards a new limit cycle. Over $N=8$ trials, the excess dissipation $\delta W_{mech}$ exhibits a sharp rise above $\Delta_{train}$, indicating the encoded memory (Fig. \ref{fig_forces}c). This result is confirmed experimentally for different values of $\Delta_{train}$ and different readout step sizes \cite{supplementary}. The protocol retrieves $\Delta_{train}$ with a precision of $2\,$mm.

Another interesting indication for the memory can be found in the statistics of force jumps in the stress-strain curves. By averaging over many realizations, we find that transients involve more events compared to limit cycles, and exhibit a different distribution of event sizes - events in the transient are typically larger (see Fig. \ref{fig_forces}d). This is also apparent when comparing the roughness of the first (blue) with the second (yellow) compression curves in Fig. \ref{fig_forces}a. Similar statistical differences were predicted to occur for irreversible vs reversible rearrangements in amorphous solids \cite{regev2021topology}. Thus, by measuring the frequency and size-distribution of stress-drops along a given driving path, an observer can, in principle, infer whether it had been visited by the system before. For example, in the experiment above, when the largest strain is exceeded the statistics and event density should change abruptly. Here, this signal is too weak to extract from a single amplitude sweep. However, we expect that in larger systems this signal will be more significant.

 We note that since the response curves are dominated by discontinuous features, the memory cannot be indicated by a peak in the bulk modulus $K=dF/d\Delta$, as done in \cite{paulsen2014multiple,chattopadhyay2022inter}. However, averaging over realizations reveals that this feature underlies the response \cite{supplementary}.

\begin{figure}[b]
\includegraphics[width=0.45\textwidth]{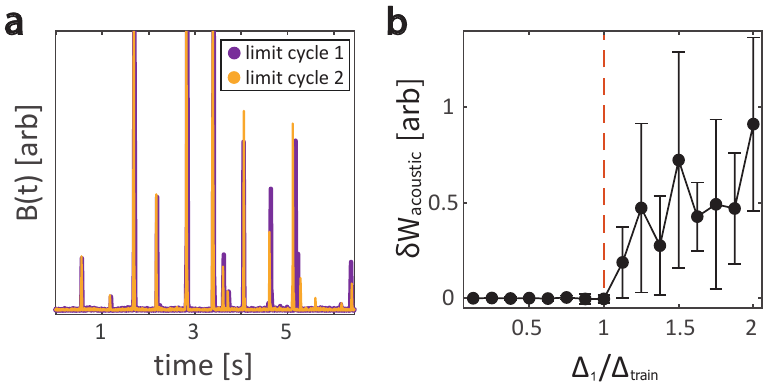}
\vspace*{-0.2cm}
\caption{\textbf{Global acoustic signal - } (\textbf{a}) Acoustic signal over time \cite{footnote_acoustic} recorded in two consecutive limit cycles after many driving cycles in a constant amplitude; (\textbf{b}) Excess emitted acoustic energy $\delta W_{acoustic}$ between two consecutive cycles, as a function of the normalized cycle amplitude $\Delta_{1} / \Delta_{train}$, averaged over $N=8$ trials. Following the same protocol as in Fig. \ref{fig_forces}b,c with $\Delta_{train}=70mm$, $\delta E_{acoustic}$ exhibits a sharp rise above $\Delta_{train}$.}
\label{fig_acoustic}
\end{figure}

\textit{Global acoustic signal - }
An alternative way to assess the irreversibly dissipated energy is by monitoring the crackling noise emitted by the crumpled sheet during the driving cycles. Whenever the sheet undergoes a local transformation, some of the energy is dissipated in the form of an acoustic pulse \cite{kramer1996universal,houle1996acoustic,shohat2022memory}. In a limit cycle, the system follows the same sequence of local transformations. Therefore, consecutive limit cycles have an identical pattern of acoustic emissions (see Fig. \ref{fig_acoustic}a) \cite{footnote_acoustic} and dissipate the same amount of energy to crackling noise. In contrast, transitory trajectories emit more pulses, and their overall dissipated acoustic energy is higher than a limit cycle for the same strain interval. Similarly to the global mechanical signal, this prototypical difference can be used to formulate an acoustic memory signal.

After a memory $\Delta_{train}$ is encoded in the sheet, we follow the protocol presented in Fig. \ref{fig_forces}c, while recording the raw acoustic signal emitted by the sheet. Namely, we oscillate the stage over $\left[\Delta_{0},\Delta_{1}\right]$, slowly increasing $\Delta_{1}$ and performing two cycles for each amplitude. The excess acoustic dissipation (in arbitrary units) for two consecutive cycles is computed as
\begin{equation}
\delta W_{acoustic}=\int\left[B_{1}^{2}(t)-B_{2}^{2}(t)\right]dt
\end{equation}
We find that, indeed $\delta W_{acoustic}$ is nearly zero for $\Delta_{1}<\Delta_{train}$, but exhibits a highly fluctuating yet prominent rise just above $\Delta_{train}$ (Fig. \ref{fig_acoustic}b), indicating excess acoustic dissipation. Interestingly, while $\delta W_{mech}$ and $\delta W_{acoustic}$ indicate the memory independently, they are only weakly correlated, with a Pearson coefficient $r=0.65$ \cite{supplementary}.

\textit{A common mechanism - }
We have presented three different signals from which the memory of the largest strain amplitude can be extracted. All three trace back to irreversible localized snapping events, which occur once the largest applied strain is surpassed. These localized changes in the geometry of the sheet underlie the sudden rise of the local signal $\sigma^2$ above $\Delta_{train}$, as shown in Fig. \ref{fig_local}d. Furthermore, each instability dissipates energy when activated. Therefore, the irreversible structural changes occurring when the maximal strain is suppressed result in excess mechanical work consumption, constituting the mechanical memory signal presented in Fig. \ref{fig_forces}c. The acoustic signal, presented in Fig. \ref{fig_acoustic}, measures a part of the same excess dissipation.

These signals are consistent with the coarse-grained description of the mechanics of crumpled sheets presented in \cite{shohat2022memory}. It was shown, that each localized instability in a crumpled sheet is a hysteretic two-level mechanical system, termed a \textit{hysteron}. Coupled to an external strain, each hysteron is defined by its flipping thresholds - the strain values in which it changes its state \cite{bense2021complex,van2021profusion}. Whenever a threshold is reached, the hysteron flips, irreversibly dissipating energy. Altogether, a crumpled sheet can be considered as a collection of interacting hysterons, where the thresholds of each hysteron depend on the state of others. Coupled hysterons models are known to reach limit cycles under cyclic drive even in the presence of interactions \cite{mungan2019networks,shohat2022memory,van2021profusion} and to hold a memory of the largest perturbation in their discrete hysteron configuration \cite{lindeman2021multiple}.

Thus, the three memory signals discussed in this work measure different signatures of the same phenomenon: excess dissipation caused by irreversible changes to the collective state of hysterons that occur whenever the system exceeds the amplitude of the limit cycle.   

\textit{Generalization - }
To demonstrate the general applicability of the global memory readout protocols, we employ them to a three-dimensional frictional amorphous material - steel wool. In this case, the microscopic structure of the steel wool is inaccessible. Moreover, the signatures of microscopic reorganization events, which are known to occur in frictional aggregates under compression \cite{weiner2020mechanics,andrade2021cohesion,bhosale2022micromechanical}, lie below the noise level of our measurements \cite{supplementary}.

\begin{figure}
\includegraphics[width=0.42\textwidth]{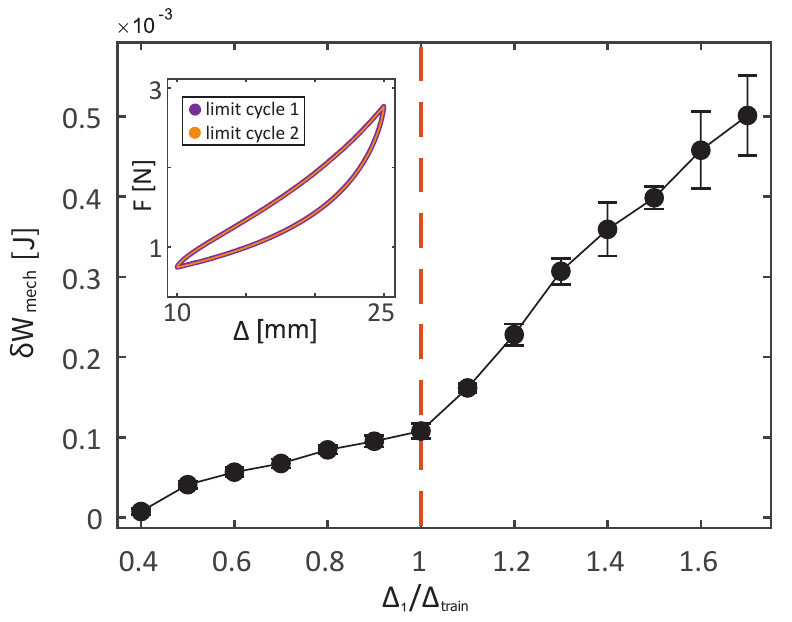}
\vspace*{-0.2cm}
\caption{\textbf{Mechanical signal in steel wool - } Excess dissipation between two cycles $\delta W_{mech}$ in steel wool, as a function of cycle amplitude $\Delta_{1} / \Delta_{train}$, for $\Delta_{train}=15\,$mm. We follow the protocol of Fig. \ref{fig_forces}b,c. Data is averaged over $N=6$ trials. (inset) Two consecutive force-displacement limit cycles, plotted after 50 training cycles.}
\label{fig_steel}
\end{figure}

We thus use the global mechanical readout protocol described above. A steel wool sample is inserted into the mechanical tester, which applies compression while measuring the force it exerts. The memory is encoded by cyclically straining the sample over the interval $\left[\Delta_{0},\Delta_{train}\right]$ until reaching an approximate limit cycle (inset of Fig. \ref{fig_steel}) \cite{supplementary}. The memory is then retrieved using the protocol described in Fig. \ref{fig_forces}b,c. We find that the global mechanical signal retrieves the trained value for $N=6$ trials, namely it exhibits a sharp rise and a kink at $\Delta_{train}$, as shown in Fig. \ref{fig_steel}. Here, below $\Delta_{train}$ the excess dissipation is non-zero. This may be due to interactions between internal degrees of freedom, resulting in slow convergence to limit cycles \cite{lindeman2021multiple}. The steel wool also exhibits a small yet measurable peak in the bulk modulus $K=dF/d\Delta$ at $\Delta_{train}$ \cite{supplementary}.

\textit{Conclusions and outlook - }
We have presented a framework for studying cyclic memory formation in complex systems, by monitoring energy dissipation during external driving. This framework links global and local indicators for the memory. Reversible cycles in the energy landscape conserve both local and global properties. Conversely, one-way transitions to new regions of the energy landscape result in irreversible local changes, accompanied by excess macroscopic energy dissipation with a clear global signature. 

We expect these global signals to be generic, and hold for a wide range of disordered systems. In particular, the local memory signatures measured in amorphous solids \cite{mukherji2019strength,keim2020global} are expected to be accompanied by excess dissipation \cite{regev2021topology}, and therefore similar global signals. This link between local and global dynamics may also facilitate the study of memory formation in systems in which the local structure is inaccessible. For example, it offers new routes to understand the mechanics of frictional aggregates and 3D amorphous solids, e.g. their slow convergence to limit cycles under periodic drive \cite{weiner2020mechanics,bhosale2022micromechanical}

Finally, this effort may help shed light on common motifs shared among different disordered systems. A wide range of these systems can be effectively described as collections of interacting (or non-interacting) hysterons \cite{bense2021complex,shohat2022memory,mungan2019networks,keim2020global,merrigan2022disorder,ding2022sequential,jules2022delicate}. Such a picture naturally captures and predicts various forms of memory formation \cite{van2021profusion,lindeman2021multiple,mungan2019structure}. However, the nature of interactions in these systems and the role they play in determining the macroscopic properties are not always clear (e.g., the observations of increased elastic moduli at the memory point \cite{supplementary,paulsen2014multiple,mukherji2019strength,chattopadhyay2022inter}). Here, the statistical differences between transients and limit cycles (Fig. \ref{fig_forces}d) are attributed to such interactions \cite{regev2021topology,shohat2022memory}. Therefore, our approach provides a global indicator for internal interactions, and may promote the understanding of their role the dynamical properties of complex and disordered systems.

\textit{Acknowledgments - } This work was supported by the Israel Science Foundation grant 2096/18. We thank Daniel Hexner, Nathan Keim, Ido Regev, Yair Shokef, and Muhittin Mungan for helpful comments.


\bibliography{bibli.bib}

\begin{thebibliography}{40}
\bibitem[{foo({\natexlab{b}})}]{footnote_acoustic}
\bibinfo{note}{We perform a Hilbert transform to the raw acoustic data, and
  display here its magnitude for better visualization}.
\end{thebibliography}

\end{document}